\documentclass[useAMS,usenatbib]{mn2e}
\usepackage[pdftex]{graphicx}

\newcommand{\beq}{\begin{equation}}
\newcommand{\eeq}{\end{equation}}
\newcommand{\beqa}{\begin{eqnarray}}
\newcommand{\eeqa}{\end{eqnarray}}

\title[Inner wind clumping]{Clumping in the inner winds of hot,
  massive stars from hydrodynamical line-driven instability simulations}

\author[Sundqvist \& Owocki]{Jon O. Sundqvist$^{1}$\thanks{E-mail:
    jon@bartol.udel.edu} and Stanley P. Owocki$^{1}$\\ $^1$University
  of Delaware, Bartol Research Institute, Newark, Delaware 19716,
  USA\\}

\begin{document}

\date{Accepted 2012 October 05. Received 2012 October 04; in original form 2012 August 14}

\pagerange{\pageref{firstpage}--\pageref{lastpage}} \pubyear{2002}

\maketitle

\label{firstpage}

\begin{abstract}

We investigate the effects of stellar limb-darkening and photospheric
perturbations for the onset of wind structure arising from the strong,
intrinsic line-deshadowing instability (LDI) of a line-driven stellar
wind. A linear perturbation analysis shows that including
limb-darkening reduces the stabilizing effect of the diffuse
radiation, leading to a net instability growth rate even at the wind
base. Numerical radiation-hydrodynamics simulations of the non-linear
evolution of this instability then show that, in comparison with
previous models assuming a uniformly bright star without base
perturbations, wind structure now develops much closer ($r \la 1.1
R_\star$) to the photosphere. This is in much better agreement with
observations of O-type stars, which typically indicate the presence of
strong clumping quite near the wind base.

\end{abstract}

\begin{keywords}
stars: early-type - stars: mass-loss - stars: winds, outflows -
hydrodynamics - instabilities
\end{keywords}

\section{Introduction}
\label{intro}

Hot, massive stars possess strong winds driven by line scattering of
the star's intense ultra-violet (UV) radiation field. The first
quantitative description of such line driving was given in the seminal
paper by \citet[][CAK]{Castor75}, who assumed a smooth, steady-state
outflow. But even though extensions of this theory
\citep[e.g.,][]{Pauldrach86, Friend86} have had considerable success
in explaining many global properties of OB-star winds, like the
predicted mass-loss dependence on metallicity and the relation between
the wind-momentum and the star's luminosity, it is nowadays clear that
these winds are in fact both highly variable and structured on a broad
range of temporal and spatial scales \citep[see][for comprehensive
  reviews]{Puls08, Sundqvist11b}.

Linear stability analyses have shown \citep[][the last two hereafter
  ORI, ORII]{Macgregor79, Owocki84, Owocki85} that the line driving of
these winds is subject to a very strong intrinsic instability,
operating on small spatial scales. And subsequent numerical modeling
of the non-linear evolution of this \textit{line-deshadowing
  instability} (LDI) has confirmed that the time-dependent wind indeed
develops a highly inhomogeneous, `clumped', structure \citep{Owocki88,
  Feldmeier95, Dessart03}.

Such structured LDI models provide a natural explanation for a number
of observed phenomena in OB-stars, such as the soft X-ray emission and
broad X-ray lines observed by orbiting telescopes like {\sc chandra}
and {\sc xmm-newton} \citep{Feldmeier97, Berghofer97, Gudel09,
    Cohen10}, the extended regions of zero residual flux typically
seen in saturated UV resonance lines \citep{Lucy83, Puls93,
  Sundqvist10}, and the migrating spectral sub-peaks superimposed on
broad optical recombination lines \citep{Eversberg98, Dessart05b,
  Lepine08}.

However, a multitude of independent observational studies also suggest
the presence of clumps in inner wind regions close to the photosphere
\citep{Eversberg98, Puls06, Sundqvist11, Najarro11, Cohen11,
  Bouret12}; this is \textit{not} reproduced by conservative,
self-excited LDI models, which develop structure only away from the
photospheric wind base \citep[at $r \ga 1.5 R_\star$,][]{Runacres02}.

This result has led to a common perception that the LDI is not able to
produce structure in inner wind regions, and so that some other
process may be the main agent responsible for the overall features of
wind clumping in OB-stars. But note that the strong damping of
structure in the inner wind found in previous instability models was a
direct consequence of a complete cancellation of the LDI by the
counteracting line-drag effect \citep{Lucy84} at the stellar surface
\citep[e.g.][hereafter OP96]{Owocki96}. But as already pointed out by
ORII, if one accounts for limb-darkening of the stellar surface
radiation, this cancellation should be incomplete and so lead to an
unstable wind base. However, whereas the effect of such
  limb-darkening has been considered before in the context of
  steady-state, line-driven winds \citep{Cranmer95, Cure12}, it has
  never been explored in time-dependent simulations.

This paper follows up on this early conjecture regarding the effect of
limb-darkening on the LDI. In \S\ref{perturb} we perform an analytic
linear stability analysis including simple Eddington limb-darkening of
the photosphere. \S\ref{numerical} then describes our numerical,
self-consistent, radiation-hydrodynamical modeling technique for
studying the non-linear evolution of the competition between the LDI
and the line-drag. \S\ref{results} examines to what extent the new
models including limb-darkening, as well as a simple photospheric
sound wave perturbation, can produce significant wind structure also
in the inner wind. Finally, \S\ref{discussion} discusses these
results, gives our conclusions, and outlines future work.

\section{Analytic perturbation analysis including limb-darkening} 
\label{perturb} 

\begin{figure}
\resizebox{\hsize}{!}{\includegraphics[]{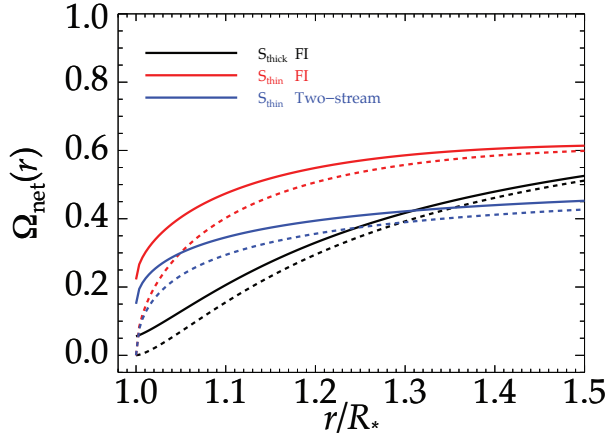}}
\caption{Relative damping effect of the line-drag on the ORI
  pure-absorption growth rate (eqn.~\ref{Eq:Om}) for
  uniformly bright (dashed lines) and limb-darkened (solid lines)
  stellar discs. Various approximations for the scattering source
  function as indicated in the figure, where `FI' denotes full angle
  integrations. Note how limb-darkened models are unstable also at the
  stellar surface ($\Omega_{\rm net}(R_\star) > 0$).}
\label{Fig:net_growth}
\end{figure}

Let us first examine the effects of limb-darkening in the linear
regime when perturbing the radiation force exerted by a single line. The
instability arises from perturbations in the direct component of the
line-force, which is proportional to the stellar core intensity
$I_{\rm c}(\mu,r) = I_\star D(\mu,r)$. Here $I_\star$ sets the scale
for the intensity and the flux-normalized disc function $D(\mu,r)$
accounts for the variation of intensity along local direction cosine
$\mu$ at radius $r$. Previous analyses have ignored limb-darkening and
simply assumed a uniformly bright stellar disc with $D=1$ for $\mu \ge
\mu_\star \equiv \sqrt{1-(R_\star/r)^2}$ and $D=0$ for $\mu <
\mu_\star$.

For a line with frequency integrated mass absorption coefficient
$\kappa$ and of `quality' $q \equiv v_{\rm th} \kappa/(\kappa_{\rm e}
c)$ \citep{Gayley95}, we can write the perturbed direct component of
the line acceleration in terms of an angle average of this core
intensity (ORI, OP96)
\begin{equation}
  \delta g_{\rm dir}(r) = \frac{4 \pi q \kappa_{\rm e}}{c} 
  \langle \mu I_\star D(r,\mu) \, \delta b(r,\mu) \rangle,  
  \label{Eq:dgdir}
\end{equation} 
where $\delta b$ is the perturbed escape probability.  For an
optically thick line with Sobolev optical depth $\tau_\mu = \kappa
\rho L_\mu \gg 1$, where $L_\mu = v_{\rm th}/(dv_{\rm n}/dn)$ is the
Sobolev length in direction $\textbf{n}$, we find for perturbations on
length scales below $L_\mu$ that $\delta b / \delta v \propto
\mu/\tau_\mu$. Upon averaging, this leads to a strong instability with
growth rate $\delta g_{\rm rad}/ \delta v \approx v/L_1$. Since this
is a factor $v/v_{\rm th}$ larger than the wind expansion rate
$dv/dr$, small initial velocity perturbations are strongly amplified
within this linear theory, by $\approx v_\infty / v_{\rm th} \approx
100$ e-folds (ORI). This implies such small-scale perturbations will
quickly reach non-linear amplitudes within pure-absorption,
non-Sobolev wind models, as first demonstrated by \citet{Owocki88}.

However, this strong de-shadowing instability can be counteracted by a
`line-drag' effect \citep{Lucy84} associated with the force of the
diffuse, scattered radiation field. Neglecting perturbations in the
source function $S$, the perturbed diffuse force term is (ORII, OP96)
\begin{equation}
   \delta g_{\rm diff}(r) = - \frac{4 \pi q \kappa_{\rm e}}{c} S(r) 
  \langle \mu \, \delta b(r,\mu) \rangle,
  \label{Eq:dgdiff}
\end{equation}   
where the minus sign signals the tendency of the line-drag to
counteract the LDI. The similarity between eqs.~\ref{Eq:dgdir} and
\ref{Eq:dgdiff} now allows us to write a very simple expression for
the net relative reduction of the pure-absorption instability
growth rate by the damping effect of this line-drag,
\begin{equation} 
 \Omega_{\rm net}(r) \equiv 
 \frac{\delta g_{\rm dir} + \delta g_{\rm diff}}{\delta g_{\rm dir}} = 
 1 - \frac{ S(r) \langle \, \mu \, \delta b(r,\mu) \, \rangle} 
 {\langle \, I_\star D(r,\mu) \, \mu \, \delta b(r,\mu) \, \rangle}.      
 \label{Eq:Om}
\end{equation}
A key issue for evaluating eqn.~\ref{Eq:Om} lies in the computation of
the assumed smooth source function $S$ for the unperturbed background
flow.  For a pure scattering line in a supersonic steady-state wind,
this can be well approximated by the local Sobolev form
\begin{equation} 
  S(r) = \frac{ \langle \, I_\star D(r,\mu) b_{\rm Sob}(r,\mu) \, \rangle}
  {\langle \, b_{\rm Sob}(r,\mu) \, \rangle},
  \label{Eq:sscat}
\end{equation} 
where the Sobolev escape probability $b_{\rm Sob}(r,\mu) = (1-\rm
e^{\it -\tau_\mu})/\tau_\mu$. For an optically thin line and a
uniformly bright disc, the source function simply follows the dilution
factor, $S(r)/I_\star = (1-\mu_\star)/2$. (Eqn.~\ref{Eq:S_ld} gives a
corresponding expression including limb-darkening.)  For a thick line,
this diluted form has a further correction for the angle-averaged
escape probability (see OP96).

\subsection{Full angle integration} 

For an optically thick line, applying eqn.~\ref{Eq:sscat} in
eqn.~\ref{Eq:Om} yields the asymptotic value $\Omega_{\rm net}
\rightarrow 0.8$ as $r \rightarrow \infty$ (ORII). Such a modest
20\,\% reduction of the pure-absorption instability growth rate
implies the wind still is extremely unstable in its outer parts.

But closer to the star the damping is stronger; indeed, for a
uniformly bright stellar disc with $S(R_\star) = I_\star/2$, the
line-drag exactly \textit{cancels} the de-shadowing instability at the
stellar surface, i.e. $\Omega_{\rm net} \rightarrow 0$ when $r
\rightarrow R_\star$.  This holds for both optically thick and thin
scattering source functions, and leads to marginal stabilization of
the wind base and to a later onset of wind structure than typically
indicated by observations.  However, noting that this exact
cancellation occurs only because $S(R_\star) = I_\star/2$, let us now
examine the effect of photospheric limb-darkening on this net growth
rate. For a simple Eddington (linear) limb-darkening law
(eqn.~\ref{Eq:ld}), analytic evaluation of eqn.~\ref{Eq:Om} gives
$\Omega_{\rm net}(R_\star) = 0.06$ and $\Omega_{\rm net}(R_\star) =
0.22$ for, respectively, an optically thick and thin source function.

This means that in such limb-darkened models the line-drag no longer 
exactly cancels the LDI at the stellar surface and thus 
that also the wind base is unstable.

\subsection{Two-stream approximation}

In a numerical wind model aiming to simulate the non-linear evolution
of the LDI, it is very computationally expensive to carry out
elaborate angle integrations at each radial grid-pint at each
time-step. But testing has shown that a simple one-ray quadrature
using a ray parameter $y \equiv (p/R_\star)^2 = 0.5$, with impact
parameter $p \equiv r \sqrt{1-\mu^2}$, actually approximates rather
well the full-angle integrated line force (see Appendix A), as long as
separate accounts are taken for inward and outward rays in the diffuse
force computations and a sphericity correction factor is applied
(OP96). For this two-stream approximation, we find
\begin{equation} 
 \Omega_{\rm net}(r) = 1 - \frac{S(r)}{I_\star} 
  \frac{2 (r/R_\star)^2}{D(\mu_y)}, 
  \label{Eq:Om_ts}
\end{equation}
where $\mu_{\rm y} \equiv \sqrt{1- y (R_\star/r)^2}$ is the local
direction cosine of the ray at radius $r$. Assuming an optically thin
source function, this again results in a zero growth rate at the
stellar surface for a uniform disc, for which eqn.~\ref{Eq:Om_ts}
actually simplifies to $\Omega_{\rm net}(r) = \mu_\star/(1+\mu_\star)$
\citep[see also][]{Owocki99}.  By contrast, for a limb-darkened disc
(again with an optically thin source function), we find
$\Omega(R_\star) = 0.15$, which is intermediate between the thin and
thick results for full angle integration.

Fig.~\ref{Fig:net_growth} plots $\Omega_{\rm net}(r)$ over $r/R_\star
= 1 - 1.5$ for the cases discussed above, assuming a smooth velocity
law $v = (1-R_\star/r)^\beta$ with $\beta =0.8$ \citep{Pauldrach86}.
Note that over this inner wind range, the two-stream approximation can
again be viewed as an intermediate case between the fully
angle-integrated cases with optically thick and thin source functions.

Most significantly, since the pure-absorption growth rate
$\delta g_{\rm dir}$ is of order 100 times the wind expansion rate,
even the base-value of $\sim$\,20\,\% of this rate found in this paper 
for limb-darkened models implies an absolute growth that is still
substantially faster, by a factor $\sim$\,20, than the wind
expansion. This suggests that non-linear structure can now develop
close to the wind base, as we next demonstrate.
 
\section{Numerical simulations of the time-dependent wind}
\label{numerical}

To follow the non-linear evolution of such instability-generated wind
structure, we must numerically integrate the radiation-hydrodynamical
conservation equations of mass, momentum, and
energy. Table~\ref{Tab:params} summarizes the stellar and wind
parameters adopted in the models, which correspond to a typical early
O-supergiant in the Galaxy. The basic simulation method used here has
been extensively described in \citet{Owocki88}, OP96,
\citet{Runacres02}; this section recapitulates key assumptions and
describes new features.

\begin{table}
	\centering
	\caption{Summary of stellar and wind parameters}
		\begin{tabular}{p{2.8cm}ll}
		\hline \hline Name & Parameter & Value \\ 
                \hline
                Stellar luminosity & $L_\star$ & 
                $8\,\times\,10^{5}\,\rm L_\odot$ \\ 
                Stellar mass & $M_\star$ & 50 \,$\rm M_\odot$ \\ 
                Stellar radius & $R_\star$ & 20 \,$\rm R_\odot$ \\
                Wind floor & & \\ 
                temperature & $T_{\rm w}$ & 40\,000\,K \\ 
                Initial steady-state & & \\
                \ - terminal speed & $v_\infty$ & 2000\,km/s \\ 
                \ - mass-loss rate & $\dot{M}$ & 
                $2.1\,\times\,10^{-6}\,\rm M_\odot/yr$ \\ 
                CAK exponent & $\alpha$ & 0.65 \\ 
                Line-strength & & \\
                \ - normalization & $\bar{Q}$ & 2000 \\
                \ - cut-off & $Q_{\rm max}$ & 0.01$\bar{Q}$ \\               
                Ratio of ion thermal & & \\
                speed to sound speed & $v_{\rm th}/a$ & 0.28 \\ 
                Electron scattering & & \\ 
                opacity & $\kappa_{\rm e}$ & 0.34\,$\rm cm^2/g$ \\ 
                \\
                \hline
		\end{tabular}
	\label{Tab:params}
\end{table}

\subsection{Radiative cooling and radiative driving} 

We solve the spherically symmetric conservation equations using the
numerical hydrodynamics code VH-1 (developed by J.~Blondin et al.).

Radiative cooling of shock-heated gas is accounted for in the energy
equation following \citet{Runacres02}. This method mimics the effect
of photoionization heating from the star's intense UV radiation field
by simply demanding that the gas never cools below a certain floor
temperature, set here to 40\,000\,K (roughly the star's effective
temperature).

A central challenge in these simulations is to compute the
line-driving in a highly structured, time-dependent wind with a
non-monotonic velocity. This requires a non-local integration of the
line-transport within each time-step of the simulation, in order to
capture the instability near and below the Sobolev length. To allow
for this objective, we adopt the smooth source function
\citep[SSF,][]{Owocki91b} method described extensively in OP96.

As in CAK, we assume a line-number distribution 
that is a power-law in line-strength $q$, but now (as in OP96) 
exponentially truncated at a maximum strength $Q_{\rm max}$,
\begin{equation}
  \frac{dN}{dq} = \frac{1}{\Gamma(\alpha) \bar{Q}} \left( \frac{q}{\bar{Q}} 
  \right)^{\alpha-2} e^{-q/Q_{\rm max}}, 
  \label{Eq:powerlaw}
\end{equation}
with $\Gamma(\alpha)$ the complete gamma function.  Here $\alpha$ is
the CAK power-law index, which can be physically interpreted as the
ratio of the line force due to optically thick lines to the total line
force, and $\bar{Q}$ is a line-strength normalization constant, which
can be interpreted as the ratio of the total line force to the
electron scattering force in the case that all lines were optically
thin\footnote{Note that we have recast the line force using the
  $\bar{Q}$ notation of \citet{Gayley95} rather than the $\kappa_0$
  notation of OP96. $\bar{Q}$ has the advantage of being a
  dimensionless measure of line-strength that is independent of the
  thermal speed. The relation between the two parameter formulations
  is $\bar{Q} = \kappa_0 v_{\rm th} \Gamma(\alpha)^{1/(1-\alpha)}/(c
  \kappa_{\rm e})$.}. For typical O-supergiant conditions at solar
metallicity, $Q_{\rm max} \approx \bar{Q} \approx 2000$
\citep{Gayley95, Puls00}. In practice, keeping the nonlinear amplitude
of the instability from exceeding the limitations of the numerical
scheme requires a significantly smaller cut-off \citep{Owocki88}.
First test-simulations \citep[see also][]{Runacres02} that
  increase $Q_{\rm max}$ to more physically realistic values indeed
  display stronger structure in wind regions where the instability is
  fully grown. But this higher $Q_{\rm max}$ also leads to spurious
  spikes in the velocity field, likely numerical artefacts caused by
  the limiting grid resolution. The full implications of the
  line-strength cut-off for wind structure is yet to be determined,
  and should be carefully examined in a detailed parameter study.
  However, we defer that to future work, since the test-simulations
  here also show that increasing the cut-off does not significantly
  affect the main focus of this paper, namely the \textit{onset} of
  non-linear structure.

All previous time-dependent wind simulations have assumed the stellar
continuum radiation can be described by a uniformly bright stellar
disc. But as discussed in \S\ref{perturb}, introducing photospheric
limb-darkening breaks the marginally stable properties of the wind
base and so may lead to structure formation also in the innermost
wind. As also noted in \S\ref{perturb}, for computational reasons we
use a simple one-ray angle quadrature for both the direct and diffuse
force components in all numerical models (see Appendix
A). Incorporating Eddington limb-darkening then typically increases
the direct line force by a modest amount, $\sim 3\,\%$, whereas the
scattering source function (and thus the diffuse line force) at the
stellar surface \textit{decreases} by $\sim 13\,\%$. This is the
essential reason there is now a net instability at the wind base.

\subsection{Boundary conditions} 

Each instability simulation evolves a smooth and relaxed Sobolev wind
model. The lower boundary is set as in \citet{Runacres02}, by fixing
the density and temperature. Previous simulations have assumed the
lowermost grid-point to be located at $r = R_\star$, the photospheric
radius where the continuum optical depth is unity, with a density
$\rho_0$ tuned to be approximately 5-10 times higher than at the sonic
point\footnote{Using a higher value leads to long-lived oscillations
  at the Lamb frequency; using lower values risk choking the
  self-regulated mass loss of the line-driven wind \citep{Owocki88,
    Owocki99}.}. However, a simple ratio estimate of the steady-state
sonic point density $\rho_{\rm a} = \dot{M}/(4 \pi a r^2)$ to the
typical photospheric density $\rho_\star \approx 1/(H \kappa_{\rm
  e})$, with scale height $H = a^2/g$, yields for the stellar and wind
parameters in table~\ref{Tab:params} a much lower value, $\rho_{\rm
  a}/\rho_\star < 0.01$. This order-of-magnitude estimate is confirmed
by more detailed, line-blanketed NLTE calculations of steady-state,
unified (photosphere+wind) model atmospheres of typical O supergiants
like $\zeta$ Pup \citep[using the {\sc fastwind} code,][]{Puls05}. In
such models, we find that the sonic point is located $\sim$\,1-2\,\%
above the stellar surface, defined by $R_\star= r(\tau_{\rm
  Ross}=2/3)$, and that the corresponding density contrast is
$\rho_{\rm a}/\rho_\star \approx 0.01$.

The time-dependent simulations reported here thus assume $r \approx
1.01 R_\star$ at the sonic point, which tends to shift inward the
stellar surface somewhat as compared to earlier models. And since the
stability properties of the lower wind are so sensitive to the
specific value of the source function, which close to the wind base
decreases very rapidly with radii, accounting for this shift can
actually be quite significant (particularly for lower-density winds).
This adds to the effect of limb-darkening in making the region near
the sonic point have a net instability.

Moreover, if one accounts also for the likelihood that the stellar
photosphere is not perfectly steady, but rather itself has a level of
variability, this can further seed the growth of unstable structure in
the overlying wind. Some previous simulations have explicitly
perturbed the lower boundary by either sound waves or turbulence
\citep[e.g.,][]{Feldmeier97}, and indeed such models tend to show
structure somewhat closer to the wind base as compared to models with
self-excited structure (compare \citealt{Runacres02} with
\citealt{Puls93} and \citealt{Sundqvist11}). To study the influence of
such perturbations in the presence of limb-darkening, the simulations
here introduce a lower boundary sound wave of amplitude $\delta
\rho/\rho_0 = 0.1$ and period 4\,000\,s.
 
\section{Numerical simulation results}
\label{results}

\subsection{Basic structure properties}

\begin{figure*}

\begin{minipage}{5.8cm}
\resizebox{\hsize}{!}{\includegraphics[]{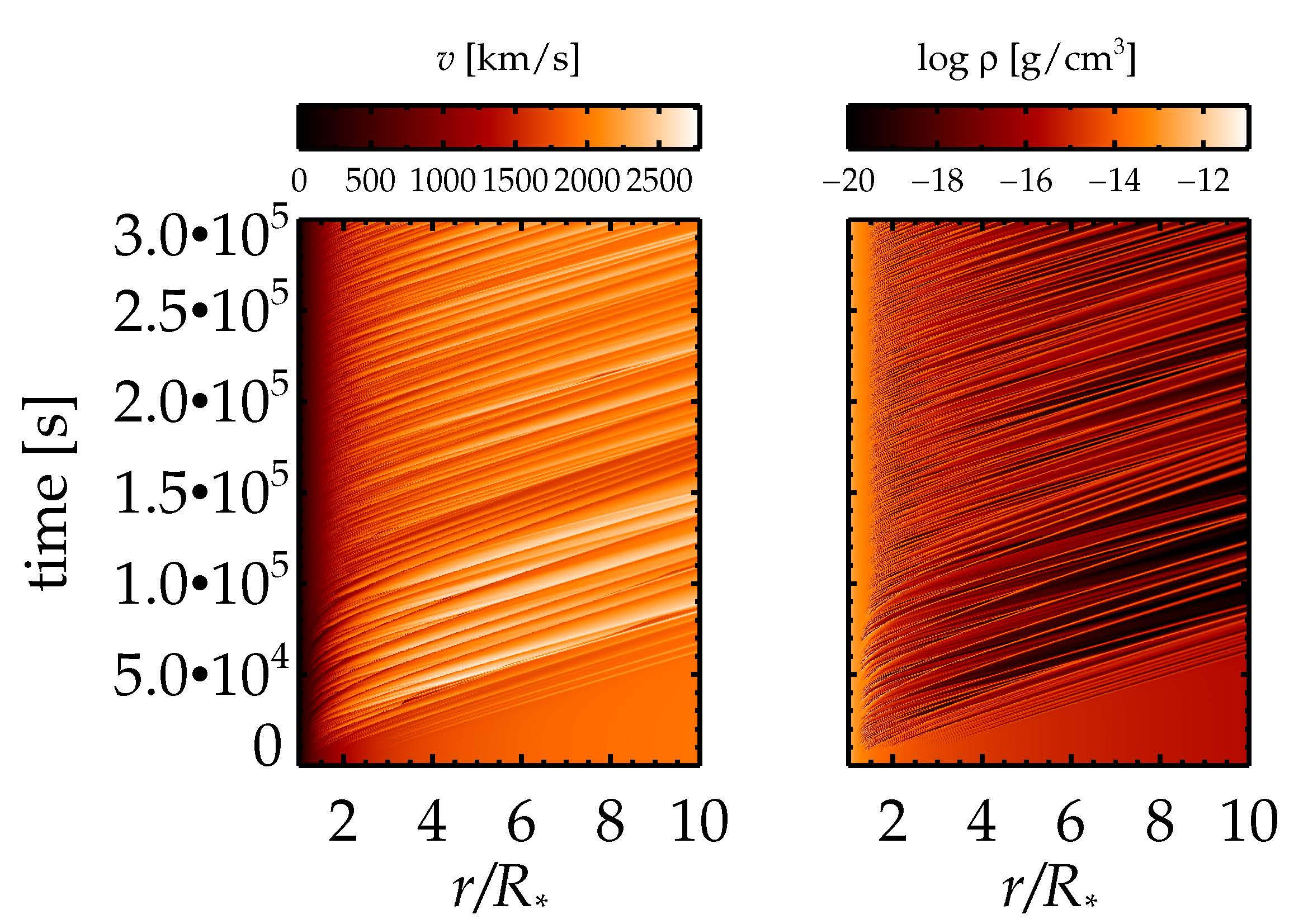}}
\end{minipage}
\begin{minipage}{5.8cm}
\resizebox{\hsize}{!}{\includegraphics[]{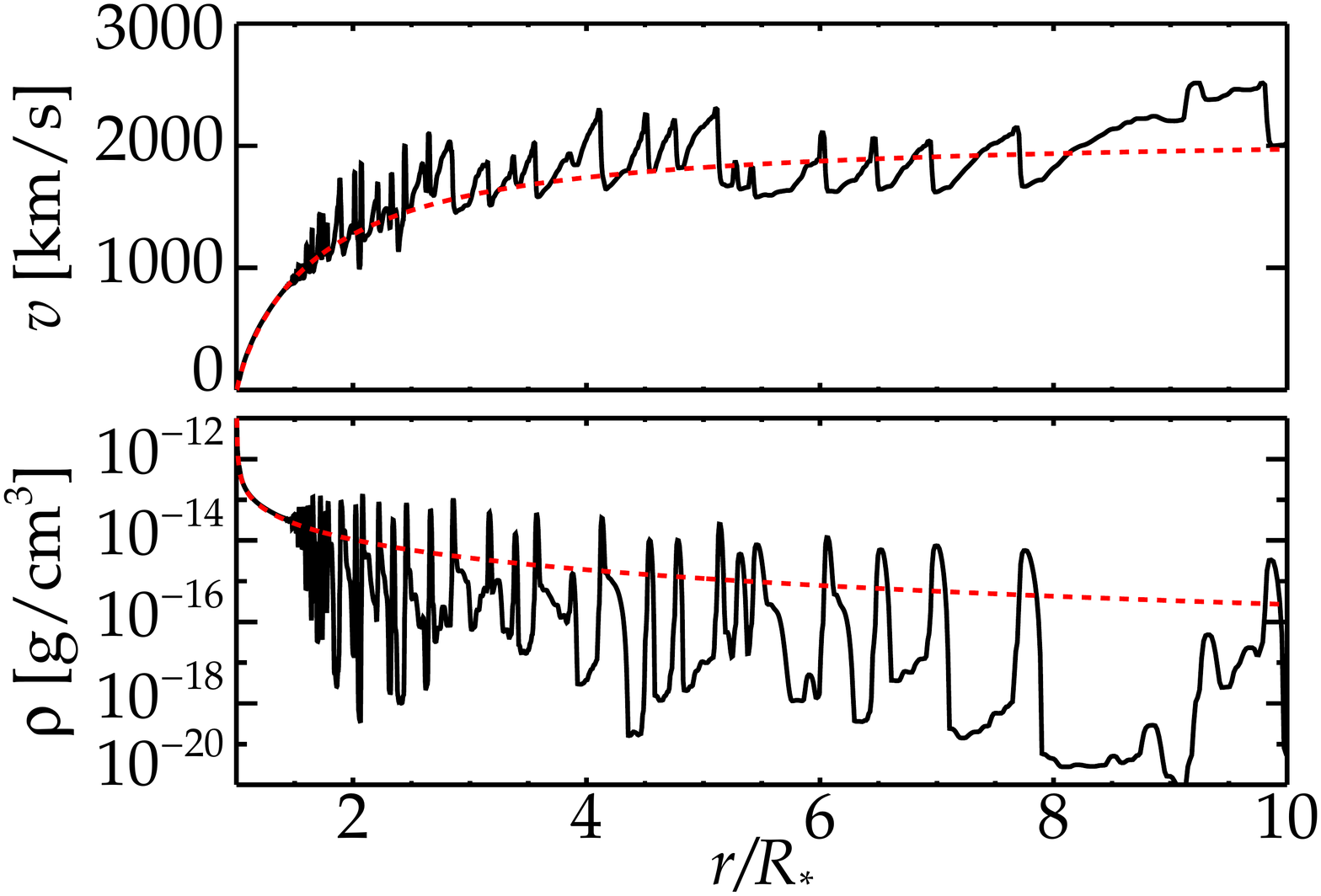}}
\end{minipage}
\begin{minipage}{5.8cm}
\resizebox{\hsize}{!}{\includegraphics[]{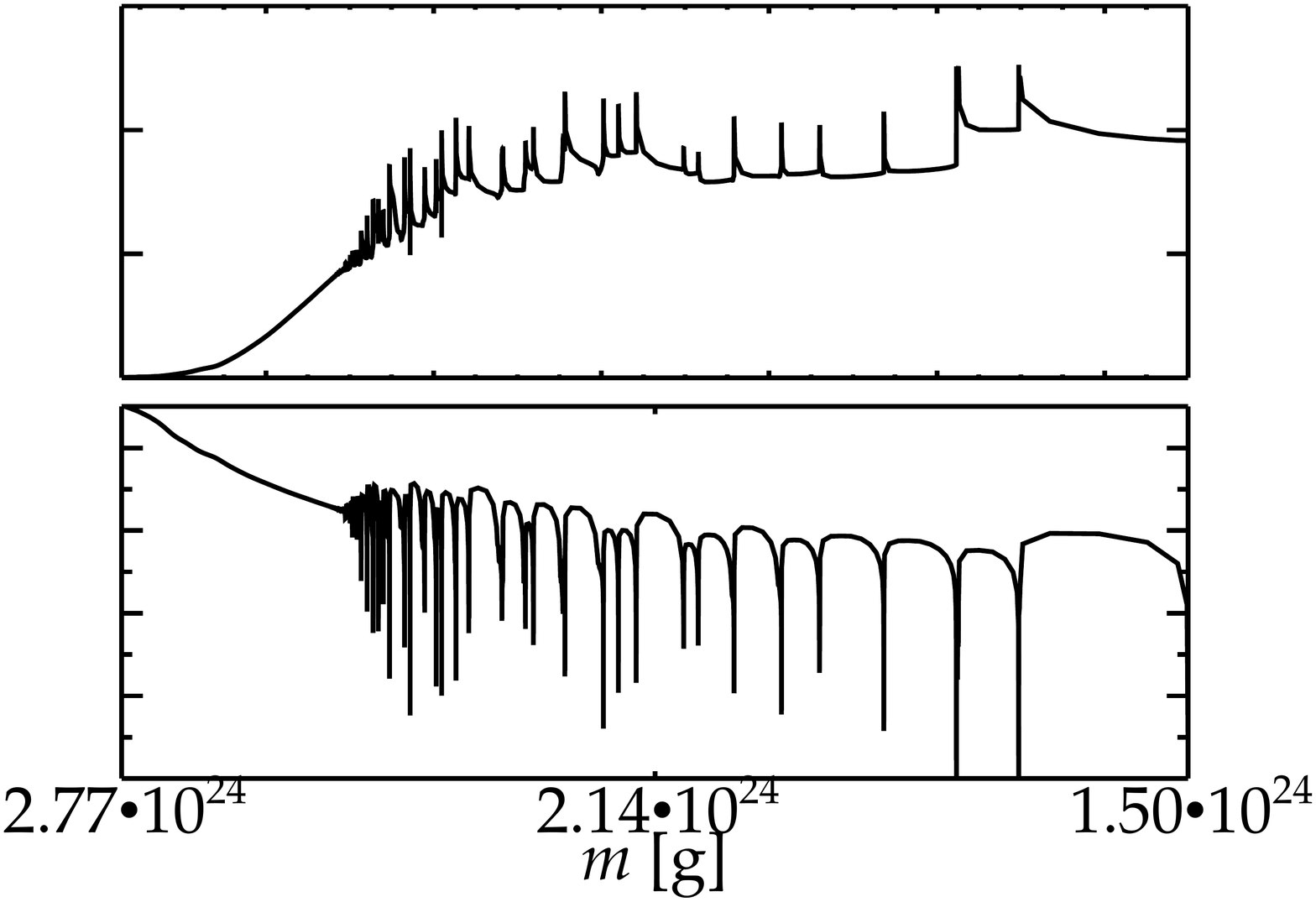}}
\end{minipage}
\caption{The left panel shows density and velocity contour plots of
  the time-evolution of a wind model with self-excited structure
  assuming a uniformly bright stellar surface. The middle and right
  panels show density and velocity of a single snapshot from the same
  model, taken $\sim 150 \, \rm ksec$ after initiation, as function of
  radius (middle) and a Lagrangian mass coordinate (right). The dashed
  red line is the smooth start model.}
\label{Fig:basic}
\end{figure*}

Let us first review the overall properties of numerical simulations
that follow the non-linear evolution of the competition between the
LDI and the line-drag \citep{Owocki99, Runacres02, Dessart03}. The
simulation displayed in Fig.~\ref{Fig:basic} illustrates the formation
of high-speed rarefactions that steepen into strong reverse shocks,
which compress most of the wind material into dense and spatially
narrow `clumps' (or really shells in these 1-D simulations). This
characteristic structure can be alternatively illustrated by plotting
density and velocity against a Lagrangian mass coordinate
tracking individual fluid elements \citep[as defined by][their
    eqn. 14]{Owocki99}; the rightmost panel of Fig.~\ref{Fig:basic}
clearly shows how the high-speed flow consists of very rarefied gas
and how most of the wind mass is concentrated into dense clumps.

This model is computed assuming a uniformly bright stellar disc and
without explicitly perturbing the lower boundary. The structure that
evolves from the smooth wind at $t = 0$ in the left panel of
Fig.~\ref{Fig:basic} is ``self-excited'', and arises from
back-scattering of radiation from outer wind structure seeding small
variations closer to the wind base, which are then subsequently
amplified by the instability \citep{Owocki99}.

Note how the line-drag effect discussed in \S\ref{perturb} greatly
suppresses instability growth close to the wind base; indeed,
structure starts to develop only at $r \approx 1.5 R_\star$ in this
simulation, at odds with observations which typically indicate the
presence of dense clumps much closer to the stellar surface (see
\S\ref{intro}). The next section examines to what extent including
limb-darkening and photospheric perturbations into the simulations can
induce wind structure also at such low radii.

\subsection{Clumping in the inner wind}

\begin{figure*}

\begin{minipage}{8.5cm}
\resizebox{\hsize}{!}{\includegraphics[]{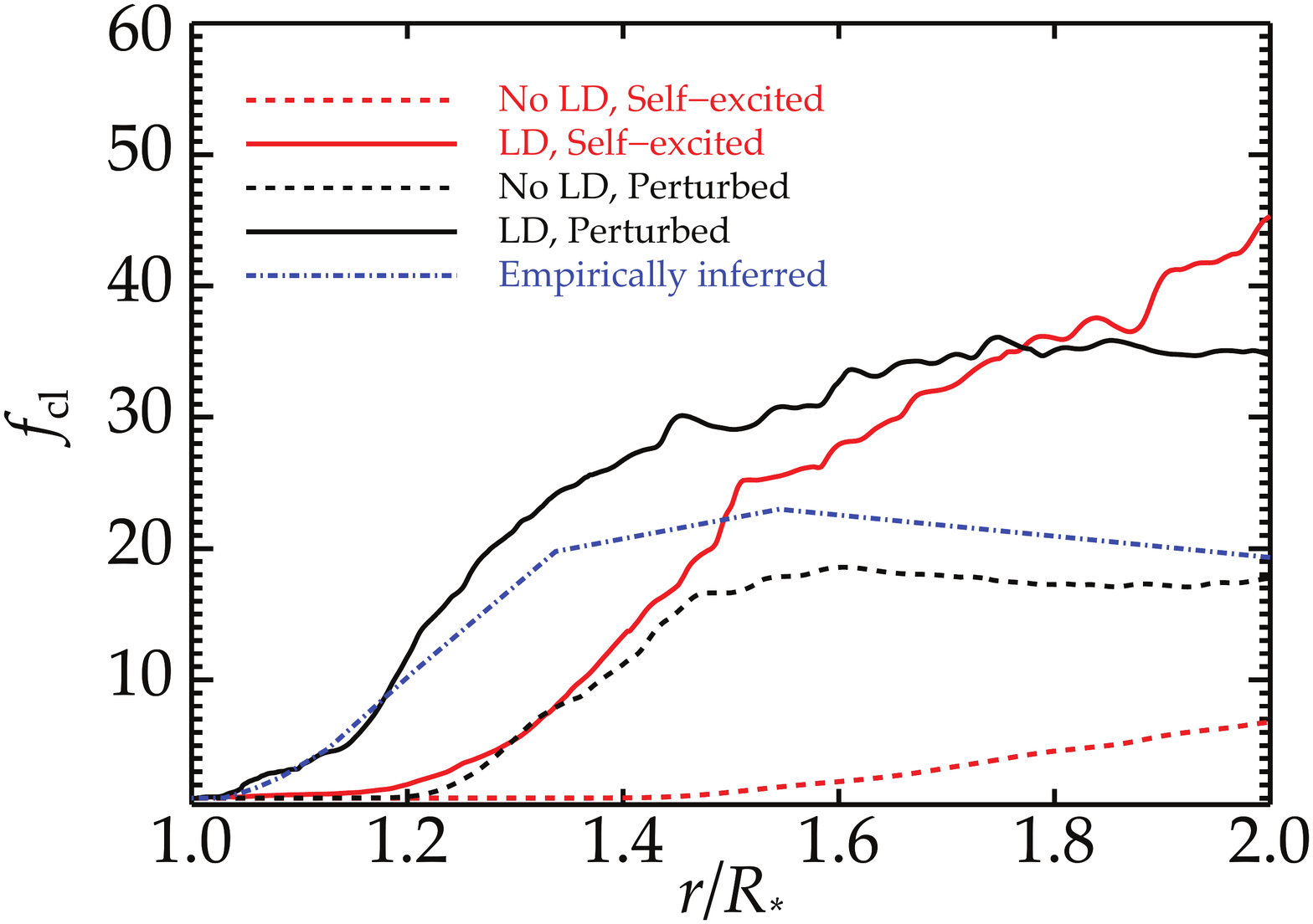}}
\end{minipage}
\begin{minipage}{8.5cm}
\resizebox{\hsize}{!}{\includegraphics[]{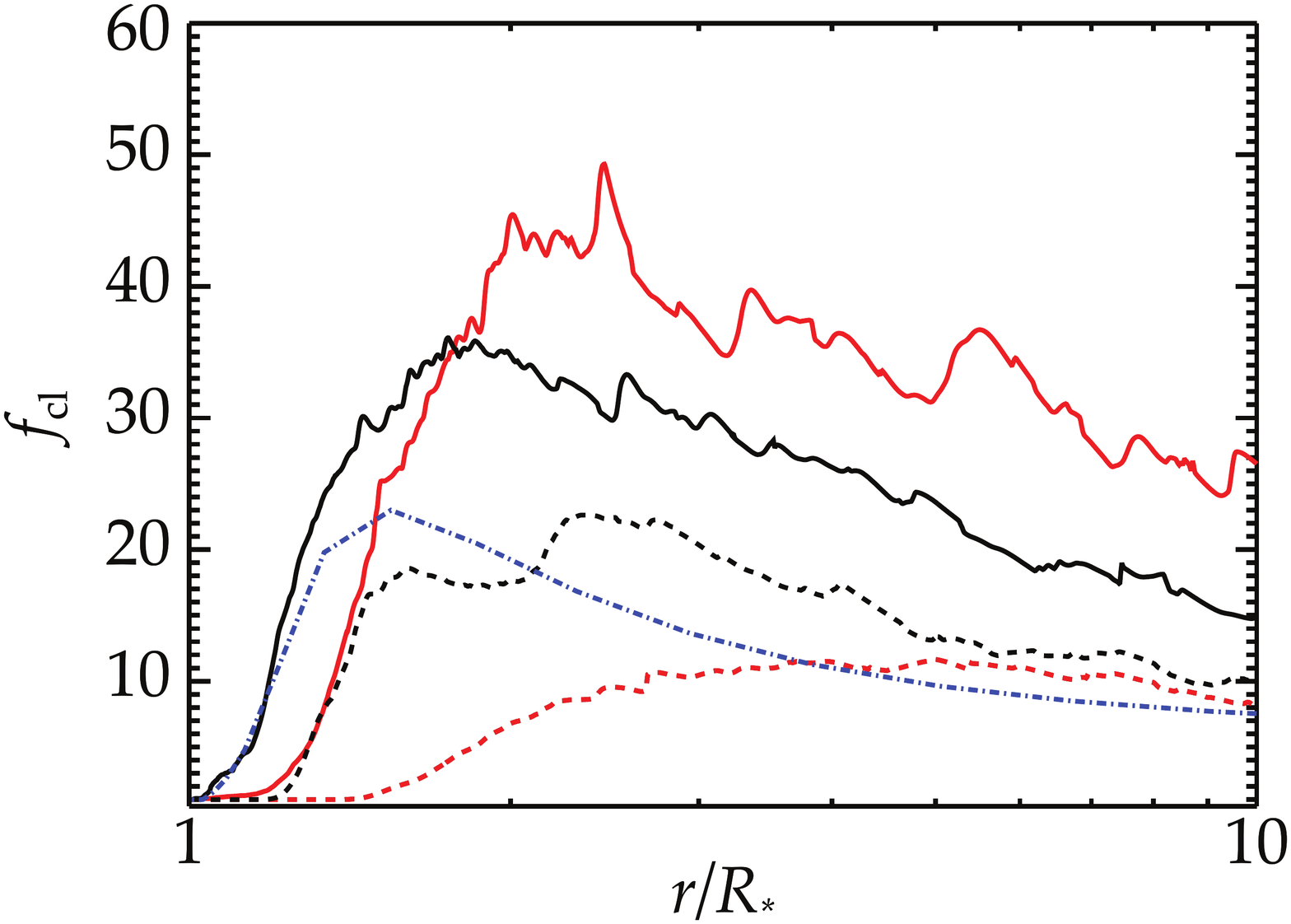}}
\end{minipage}
\caption{Simulated clumping factors $f_{\rm cl}(r)$. The left panel
  focuses on the inner wind regions $r/R_\star = 1 - 2$, whereas the
  right panel shows the wind all the way out to $r/R_\star = 10$ on a
  logarithmic abscissa. Blue dashed-dotted lines compare the
  simulations to the clumping law inferred for $\zeta$ Pup by
  \citet{Najarro11}, see text. `LD' indicates whether limb-darkening
  is accounted for in the models.}
\label{Fig:fcl}

\end{figure*}

\begin{figure*}

\begin{minipage}{8.5cm}
\resizebox{\hsize}{!}{\includegraphics[]{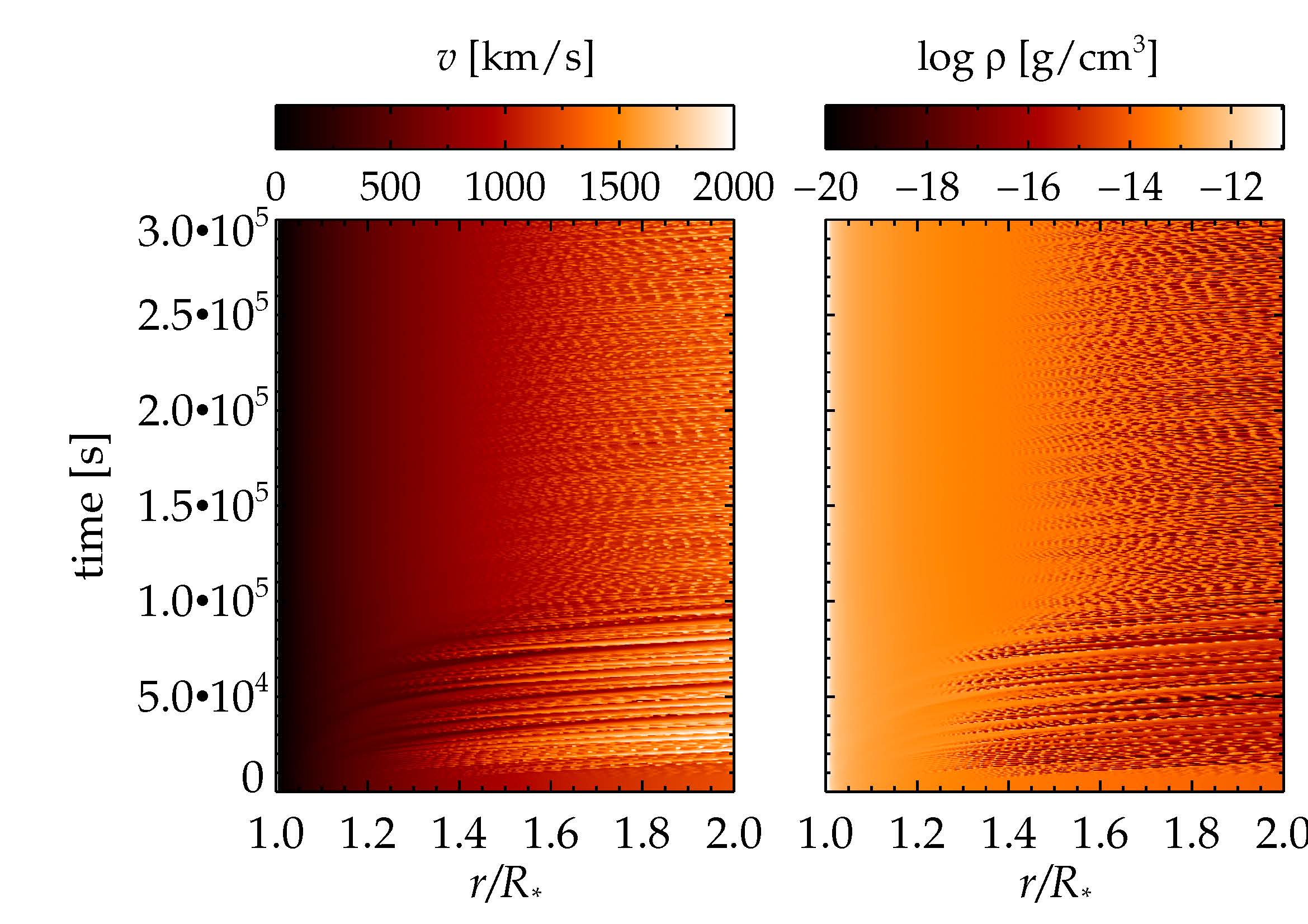}}
\end{minipage}
\begin{minipage}{8.5cm}
\resizebox{\hsize}{!}{\includegraphics[]{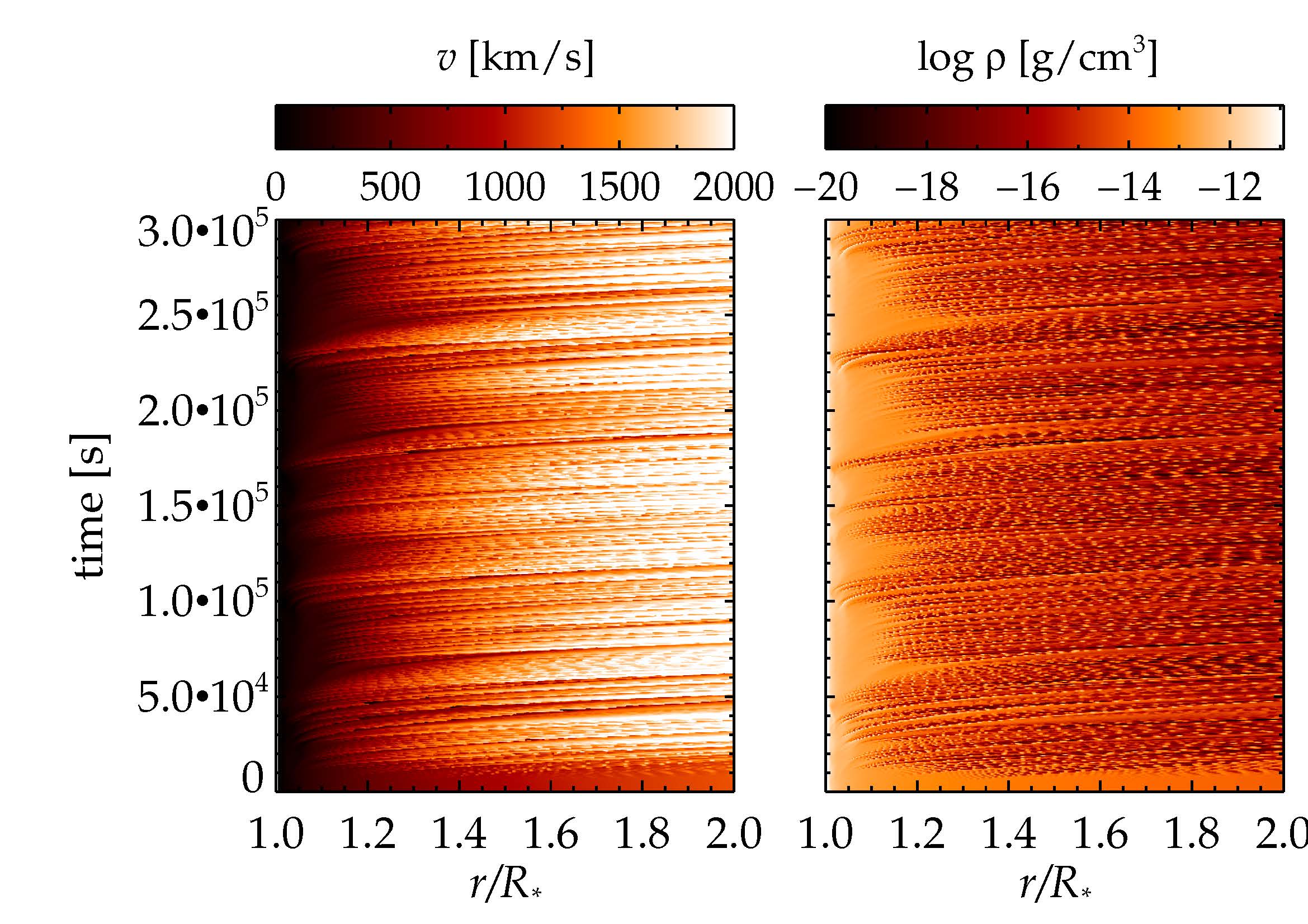}}
\end{minipage}
\caption{Inner wind time evolutions of a simulation without
    limb-darkening and photospheric perturbations (left) and one
    including both effects (right).}
\label{Fig:time_ev}

\end{figure*}

Let us characterize the wind density structure in terms of a clumping factor  
\begin{equation}
  f_{\rm cl} = \langle \rho^2 \rangle / \langle \rho \rangle^2,
\end{equation}
where the angle brackets here denote time
averaging. Fig.~\ref{Fig:fcl} plots the radial variation of such
clumping factors for the four simulation cases considered, i.e. for
uniform-disc and limb-darkened models with and without explicit base
perturbations, computed by averaging over each simulation time-step
between $t=100-300 \, \rm ksec$, where the simulations have become
insensitive to the initial conditions.

The right panel shows $f_{\rm cl}$ over the full simulation range
$r/R_\star = 1-10$, whereas the left panel focuses on the crucial
inner wind regions where limb-darkening and base perturbations alter
the onset of wind structure. The blue dashed curve in both panels
show, for comparison, the empirically inferred clumping
factor\footnote{In such empirical work, $f_{\rm cl}$ is typically
  defined using volume-averaging rather than time-averaging, but in
  the stochastic medium here the two should be effectively
  interchangeable. If one neglects the small inter-clump density
  \citep[but see][]{Zsargo08, Sundqvist10, Surlan12}, this clumping
  factor is equal to the inverse of the clump volume filling factor,
  $f_{\rm cl} = 1/f_{\rm vol}$.} for $\zeta$ Pup based on the
comprehensive multi-diagnostic study by \citet{Najarro11}.

The left panel illustrates that both perturbations and limb-darkening
shift inward the onset of clumping as compared to the standard model.
Much as anticipated in the linear stability analysis, limb-darkening
increases the net growth rate, while the perturbation provides a seed
which the instability can amplify to form the clumped structure. The
right panel shows that this stronger structure persists to several
stellar radii from the wind base. But note that our simulation volume
here only extends to $ \approx 10 R_\star$; the analysis of
\citet{Runacres02} suggests that such strong structure will eventually
dissipate and tend to settle at $f_{\rm cl} \approx 4$ in the
outermost, radio emitting wind.

We emphasize here that this initial exploration study only attempts to
outline the general importance of limb-darkening and perturbations for
the onset of LDI generated wind structure. Thus we do not aim for a
perfect match to the empirical $\zeta$ Pup clumping law throughout the
wind; quantitative details in the predicted clumping factors will
depend on the exact treatment of limb-darkening and the source
function, the nature and strength of the applied base perturbations,
and the level of lateral fragmentation of clumps in multi-dimensional
LDI simulations \citep{Dessart03, Dessart05}, as further discussed in
\S\ref{discussion}. Nonetheless, the new overall agreement between
simulations and observations in the inner wind is encouraging, and
suggests that the standard line-deshadowing instability
(possibly seeded by some modest photospheric perturbations) is
indeed fully capable of generating significant structure also near
the wind base.

\subsection{Time-evolution with and without limb-darkening and base perturbations}

Fig.~\ref{Fig:time_ev} compares the time evolution of velocity and
density in the inner wind for the model including limb-darkening and
base perturbations (right) against the model ignoring both effects
(left). Following the brief adjustment to initial conditions, the
unperturbed model settles to a state with onset of intrinsic structure
and variability at $r \approx 1.5 R_\star$.  In contrast, the
perturbed, limb-darkened model shows a much earlier onset that varies
from a maximum $r \approx 1.15 R_\star$ all the way down to the
photospheric wind base.  The characteristic time-scale of this
variation, $\sim$\,50\,ksec, is much longer than the perturbation
period of $\sim$\,4\,ksec.

Our preliminary analysis suggests that this long-term variation is
likely related to the intrinsic variations found in the
``pure-absorption'' models of \citet{Poe90}. As discussed there, these
variations stem from a degeneracy of the CAK-like steady-state
solutions in the regions near the wind sonic point. As shown in
Fig.~7b of \citet{Poe90}, this leads to a wind-speed oscillation with
period $\sim$\,50\,ksec. For slightly different conditions, the
oscillating flow can actually even stagnate and re-accrete onto the
star \citep[see Fig.~1 in][]{Owocki91}.

Remarkably, in the standard SSF model the introduction of the diffuse
force component eliminates this solution degeneracy, and so suppresses
the oscillatory behaviour. But the additional introduction here of
limb-darkening now enhances the direct pure-absorption component of
the line-force, and reduces the stabilizing diffuse component. This
evidently allows the degeneracy to reappear, and so again leads to
slow variations in the wind solution. The full consequences of this
effect for wind structure and variability (for example, as a
possible origin of discrete absorption components,
\citealt{Howarth89}) must await multi-dimensional LDI wind models
\citep{Dessart03, Dessart05}. Similarly, implications for wind
initiation must await a future, more complete analysis of the
dynamical role of diffuse radiation near the sub-sonic wind base
\citep[see, e.g.,][]{Owocki99}. Note though, that in the current
models the mass flux initiated at the wind base is still quite close
to the initial steady state derived from standard CAK Sobolev theory.

\section{discussion, conclusions, and future work}
\label{discussion}

The central result of this paper is that photospheric limb-darkening
in instability wind models leads to structure growth closer to the
wind base than in previous models assuming a uniformly bright stellar
disc. We demonstrate this both analytically in a linear perturbation
analysis, and numerically using self-consistent, time-dependent
radiation-hydrodynamical wind simulations. Particularly when combined
with perturbations of the lower boundary, such limb-darkened models
reproduce well the early onset of clumping typically inferred from
observations of O-star winds.

The analysis here uses a simple sound wave to examine the general
effects of lower boundary perturbations for the onset of LDI-generated
wind structure. Physically, photospheric perturbations could, for
example, originate from non-radial pulsations, or even from the thin
sub-surface convective layer associated with the iron opacity peak
\citep{Cantiello09}.

Moreover, as in previous models, the mass flux at the wind base in
these time-dependent simulations follows quite closely that of the
initial steady state derived from standard CAK theory, This suggests
the early onset of wind structure found here does not directly alter
the average mass-loss rate of the star. But note in this respect that
the resulting clumped structure will modify e.g. the wind ionization
balance, and so \textit{indirectly} affect the line-force. Such
feedback effects are not included in the simulations presented here,
but may impact theoretical mass-loss rates derived from steady-state
models \citep{Muijres11}.

For simplicity the simulations here assume spherical symmetry; first
2-D models by \citet{Dessart03, Dessart05} suggest somewhat lower
clumping factors than in comparable 1-D models, by a factor of
approximately two. Future work should carry out such multi-D LDI
simulations including also the effects of limb-darkening and base
perturbations.

Finally, an unanticipated result of this study is the re-appearance of
the solution degeneracy and associated slow global wind variations
found in the pure-absorption models of \citet{Poe90}. This arises from
the change in the relative strength of the diffuse vs. direct force
components in the transonic wind region of limb-darkened
simulations. Determining the broad implications of these effects for
wind structure and variability will require a careful analysis of the
diffuse radiation in this wind initiation region around the sonic
point. Thus, although standard CAK wind models focus solely on the
direct radiation from the star, this again \citep[see also,
  e.g.,][]{Owocki99} emphasizes the subtle role of diffuse scattered
radiation in the dynamics of line-driven winds.
 
\section*{Acknowledgments}

This work was supported in part by NASA ATP grant NNX11AC40G.
We would also like to thank the referee for useful comments on
the manuscript.

\newpage 

\appendix 

\section{The effect of stellar limb-darkening on the radiation line force}

\subsection{The direct force term} 

The direct line force term from an ensemble of lines at a distance $r$
from the stellar center in spherical symmetry may be written as

\begin{equation} 
  g_{\rm dir}(r) = g_{\rm e} \bar{Q} \,
  \frac{\langle \, \mu  D(\mu,r) b(\mu,r) \, \rangle}
  {\langle \, \mu  D(\mu,r) \, \rangle},   
  \label{Eq:dir}
\end{equation}
which is just OP96's eqn.~58 recast in \citeauthor{Gayley95}'s (1995)
$\bar{Q}$ notation. Here $g_{\rm e} \bar{Q} \equiv g_{\rm thin}$ is
the line force if all lines were optically thin and
\begin{equation} 
  b(\mu,r) = \int_{-\infty}^{\infty} \frac{\phi[\,x - \mu v/v_{\rm th}\,]}
  {[\,t_{\rm \bar{Q}}(x,\mu,r,z_{\rm b}) + \bar{Q}/Q_{\rm max}\,]^\alpha} dx
  \label{Eq:besc}
\end{equation} 
the \textit{ensemble-summed} escape probability (OP96, eqn.~57), where
$\phi$ is the profile-function (assumed here to be a Gaussian) and
$t_{\rm \bar{Q}}(x,\mu,r,z_{\rm b})$ the frequency-dependent optical
depth for a line of strength $q = \bar{Q}$.

Assuming a uniformly bright stellar disc, and casting the angle
integration in terms of a ray parameter $y \equiv (p/R_\star)^2$ (see
\S\ref{perturb}), we may write eqn.~\ref{Eq:dir} as
\begin{equation} 
  g_{\rm dir}(r) = g_{\rm thin} \int_0^1 b(\mu_{\rm y},r) dy,  
  \label{Eq:dir_y}
\end{equation}
with $\mu_{\rm y}(r) = \sqrt{1-y(R_\star/r)^2}$ the direction cosine
of the ray.

Let us next consider a linear Eddington limb-darkening law \citep[see,
  e.g.,][eqn. 3.37]{Mihalas78},
\begin{eqnarray}
  D(\mu,r) &=& \frac{1}{2} + \frac{3}{4} \mu' = \frac{1}{2}
  + \frac{3}{4} \sqrt{\frac{\mu^2-\mu_\star^2}{1-\mu_\star^2}} \nonumber \\
  &=&\frac{1}{2} + \frac{3}{4} \sqrt{1-y},
   \label{Eq:ld}  
\end{eqnarray} 
where $\mu' = \cos \theta'$ is the direction cosine between a
\textit{star-centered} radius vector and the ray at the stellar
surface, and where $\mu_\star$ and $\mu$ are related to,
respectively, the dilution factor and impact parameter (as defined
in \S\ref{perturb}).

\begin{figure}
\centering
\resizebox{\hsize}{!}{\includegraphics[]{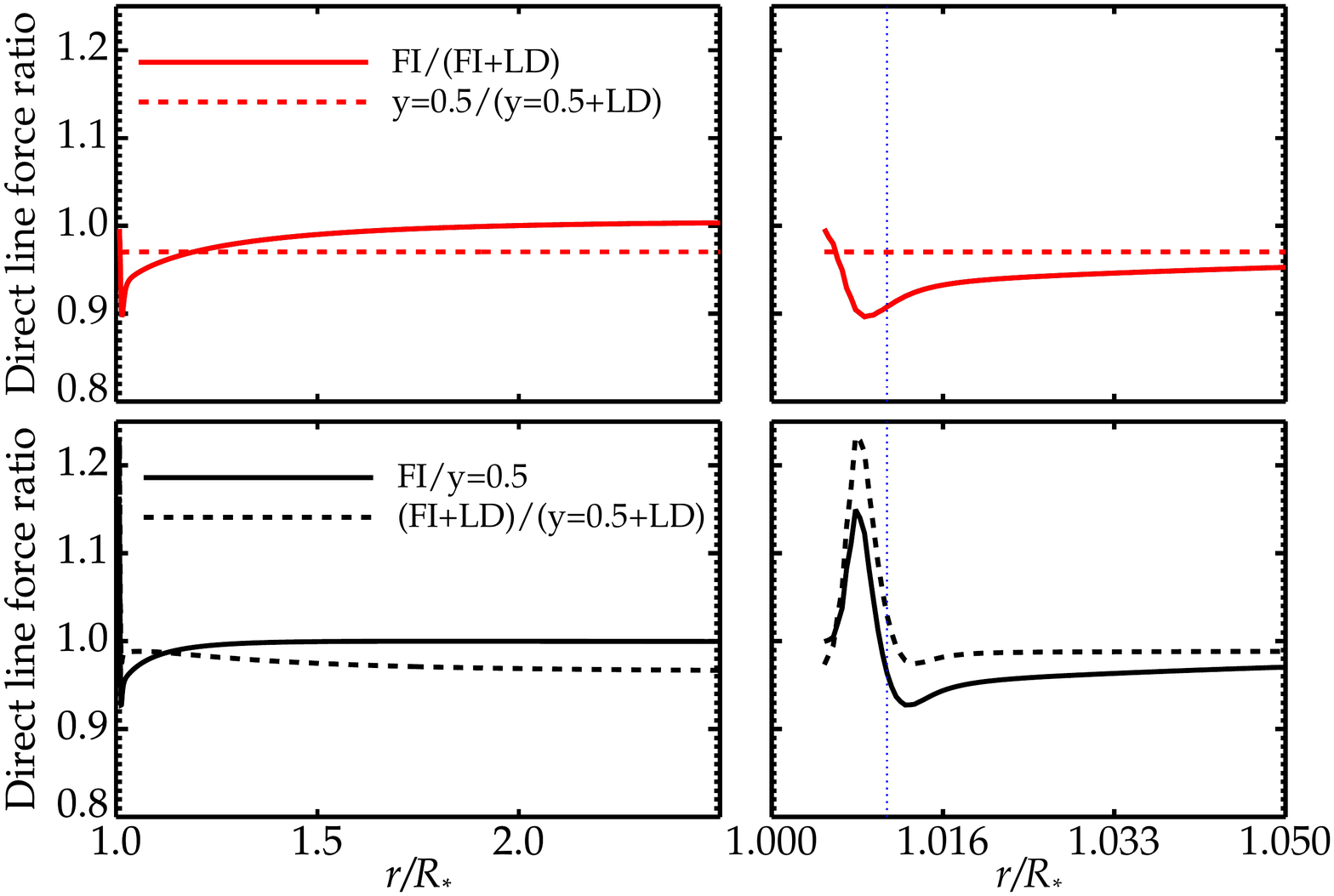}}
\caption{Upper panels show ratios of direct line-forces computed with
  and without limb-darkening (LD), for full angle-integrations (FI)
  and a one-ray quadrature with $y=0.5$. Lower panels show ratios of
  forces computed with full angle-integrations and the same one-ray
  quadrature, for uniform-disc and limb-darkened models. The right
  panels focuses on the innermost wind, with the blue dotted lines
  marking the position of the steady-state sonic point.}
\label{Fig:force}
\end{figure}

Applying this limb-darkening law in eqn.~\ref{Eq:dir}
yields
\begin{eqnarray} 
  &&g^{\rm ld}_{\rm dir}(r) = g_{\rm thin} \times \nonumber \\ 
  && \left(\frac{1}{2}\int_0^1 b(\mu_{\rm y},r) dy +   
  +\frac{3}{4} \int_0^1 
  \sqrt{1-y} \, b(\mu_{\rm y},r) dy \right). 
  \label{Eq:dir_ld}
\end{eqnarray}
In a time-dependent wind model, these angle integrals must be
evaluated numerically. However, to carry out elaborate angle
integrations at each radial grid-pint at each time-step is very
computationally expensive. Fig.~\ref{Fig:force} plots line forces
computed at the first evolution time-step of a previously relaxed
Sobolev model, using i) full numerical angle-integrations of
eqn.~\ref{Eq:dir_y} and \ref{Eq:dir_ld} and ii) a simple one-ray
quadrature with $y = 0.5$. The figure shows that for both
limb-darkened and uniform-disc models this one-ray quadrature gives
quite accurate results, within 10-20\,\% of the full angle integration
in the sub-sonic region and much better in the part of the wind where
structure develops. As in previous uniform-disc LDI simulations
\citep[e.g., OP96;][]{Owocki99, Runacres02}, all time-dependent wind
models presented in the main text of this paper use this one-ray
formulation.

Note further from Fig.~\ref{Fig:force} that nowhere in the wind is the
effect of limb-darkening on the direct line force bigger than
$\sim$\,10\,\%.
  
\subsection{The diffuse force term} 

The diffuse line force in the smooth source function (SSF)
approximation is (OP96, eqn 62)
\begin{equation} 
  g_{\rm diff}(r) = - 2 g_{\rm thin} \frac{S(r)}{I_\star} \,
  \langle \, \mu b(\mu,r) \, \rangle.   
  \label{Eq:diff}
\end{equation}
Using again the ray parameter $y$, we may approximate
eqn.~\ref{Eq:diff} by
\begin{eqnarray} 
	g_{\rm diff}(r) &\approx& \frac{S(r)}{S_{\rm
            t}(r)}\frac{g_{\rm thin}}{2(1+\mu_\star)} \times 
        \nonumber \\ && \int_0^{1}
        (b_-(\mu_{\rm y},r) - b_+(\mu_{\rm y},r)) dy,
	\label{Eq:diff_y}
\end{eqnarray}	 
where we have now taken separate accounts for outward streaming
(``$+$'', $\mu > 0$) and inward streaming (``$-$'', $\mu < 0$)
photons, and applied a simple ($r/R_\star)^2$ correction factor to
account for the fact that the angle integral here really should extend
to $y=(r/R_\star)^2$ \citep[][OP96]{Owocki91b}. Expressions for the
corresponding optical depths can be found in OP96.

$S_{\rm t}(r)/I_\star = (1-\mu_\star)/2$ is here the optically thin,
pure scattering source function in the case of a uniformly bright
stellar disc (eqn.~\ref{Eq:sscat}).  Applying the Eddington
limb-darkening law eqn.~\ref{Eq:ld} in eqn.~\ref{Eq:sscat}, this
optically thin source function now takes the form
\begin{equation} 
  \frac{S^{\rm ld}_{\rm t}(r)}{I_\star} = \frac{1}{16} 
  \left(7-4 \mu_\star + 3 \mu_\star^2 \, \frac{\ln   	
  [\mu_\star/(1+\sqrt{1-\mu_\star^2})]}{\sqrt{1-\mu_\star^2}}\right), 
  \label{Eq:S_ld}   
\end{equation}
which takes the value 7/16 (i.e. $<$\,1/2) at the stellar surface,
where $\mu_\star \rightarrow 0$. By simply applying this source
function in eqn.~\ref{Eq:diff_y}, we can then approximate the effect
of stellar limb-darkening on the diffuse line-force component.
   
\subsection{Sobolev approximation}

\begin{figure}
\resizebox{\hsize}{!}{\includegraphics[]{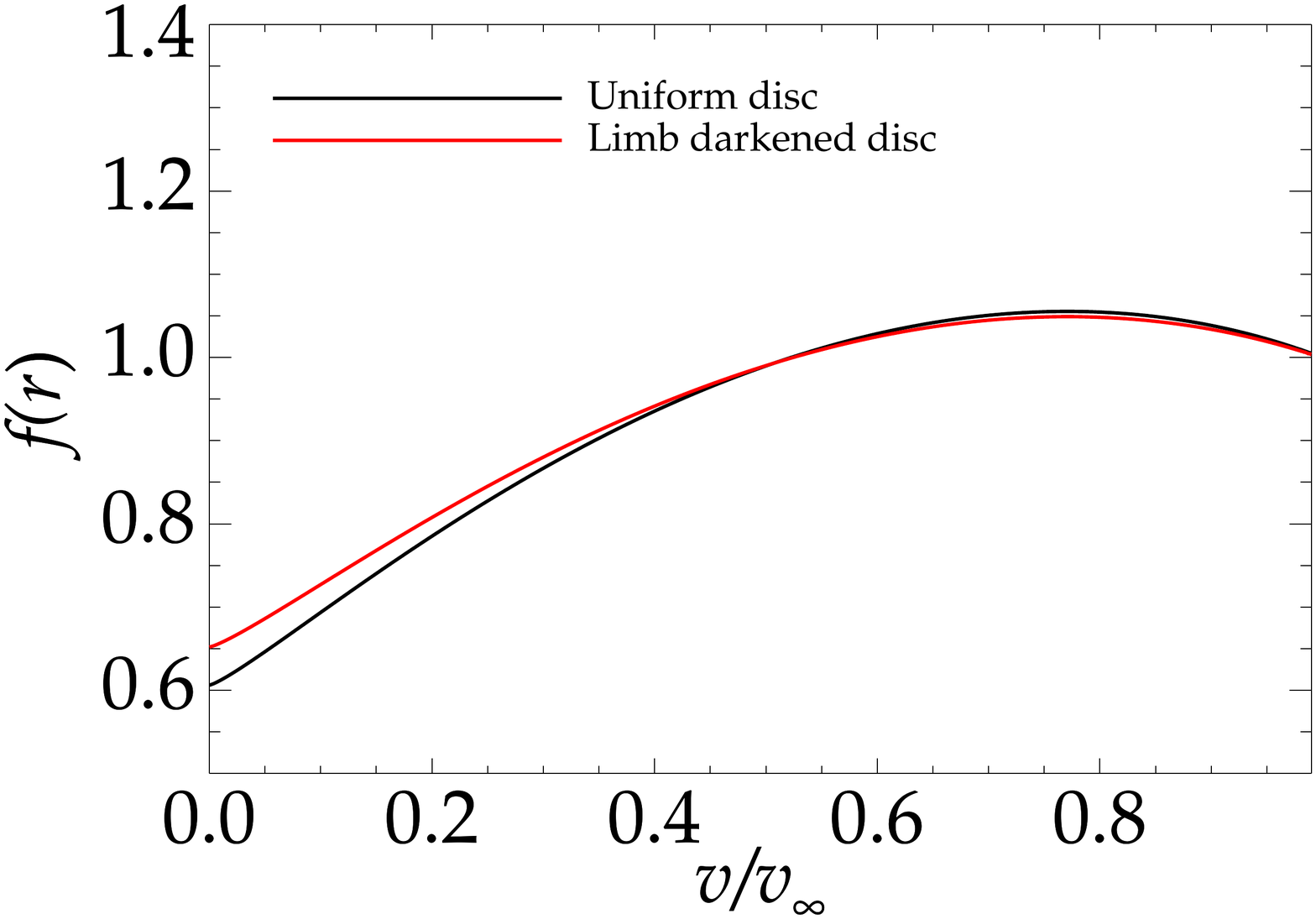}}
\caption{Finite-disc correction factors for a uniform (black) and
  limb-darkened (red) disc, assuming a smooth $\beta=0.8$ velocity law
  and $\alpha=0.65$.}
\label{Fig:fd_fac}
\end{figure} 

For completeness, we add corresponding limb-darkening modifications
also for the Sobolev line force. Now
eqn.~\ref{Eq:besc} can be solved analytically to yield for the CAK
direct line force (which is also the \textit{total} line force since
in the Sobolev approximation the diffuse force vanishes everywhere)
\begin{equation}
  g_{\rm dir}^{\rm Sob} = \frac{g_{\rm thin}}{(1-\alpha)
    \tau^\alpha} f(r) \left( \frac{1+\tau_{\rm max}}{\tau_{\rm
      max}}^{1-\alpha}- \frac{1}{\tau_{\rm max}}^{1-\alpha} \right),
  \label{Eq:dir_sob}
\end{equation}
with Sobolev optical depths $\tau = \bar{Q} t \equiv \bar{Q}
\kappa_{\rm e} \rho c/(dv/dr)$ and $\tau_{\rm max} = Q_{\rm max} t$,
and finite-disc correction factor
\begin{eqnarray} 
  f(r) \equiv \frac{g_{\rm Sob}^{\rm fd}}{g_{\rm Sob}^{\rm rad}} &=&
    \frac{2}{(1-\mu_\star^2)(1+\sigma)^\alpha } \\
    &&\int_{\mu_\star}^1 D(\mu,r) \mu (1+\sigma \mu^2)^\alpha d \mu, 
    \label{Eq:fd} 
\end{eqnarray}
where $\sigma \equiv \rm d(\it \ln v)/\rm d(\it \ln r) \rm - 1$ is the
so-called wind anisotropy factor.

For the case of a uniformly bright disc with $D=1$, eqn.~\ref{Eq:fd} becomes
\begin{equation} 
  f(r) = \frac{(1+\sigma)^{1+\alpha} - (1+\sigma \mu_\star^2)^{1+\alpha}}
  {(1+\alpha)\sigma(1+\sigma)^\alpha (1-\mu_\star^2)}. 
  \label{Eq:fd_nold}
\end{equation} 
As $r \rightarrow \infty$, $\mu_\star \rightarrow 1$ and $\sigma
\rightarrow - 1$, and thus $f \rightarrow 1$, as expected since at
large distances the star closely resembles a point source. But closer
to the star, where $\sigma > 1$, the finite-disc correction reduces
the line-force somewhat, with the limiting value $f(r \rightarrow
R_\star) \rightarrow 1/(1+\alpha)$.

For a limb-darkened disc, eqn.~\ref{Eq:fd} cannot generally
be solved analytically. However, again taking the near-star limit 
for the simple Eddington-case, one obtains \citep{Cranmer95}
\begin{equation} 
  f^{\rm ld}(R_\star) = \frac{1}{1+\alpha} \left[ \frac{1}{2} +
    \frac{3}{2} \left( \frac{1+ \alpha}{3 + 2 \alpha} \right) \right].
    \label{Eq:fd_ld}
\end{equation}
The square brackets are here the limb-darkening correction factor,
equal to unity for optically thin lines with $\alpha=0$, but always
somewhat greater than unity (up to 1.2 for $\alpha=1$) for a realistic
line-ensemble with optically thick line-fraction $\alpha > 0$, simply
due to the more centrally concentrated stellar light.

Fig.~\ref{Fig:fd_fac} compares the two radial finite-disc functions
for $\alpha=0.65$ and a canonical $v = v_\infty (1-R_\star/r)^\beta$
velocity law, using $\beta=0.8$. This again shows that the general
effect of limb-darkening is modest, with the minimum ratio $f/f^{\rm
  ld} \approx 0.93$ occurring at the stellar surface. Such quite small
effects are in accordance with the results found in the previous
section for non-Sobolev line forces.

Indeed, from numerical hydrodynamics simulations that relax a
  Sobolev line-force model, including full angle-integrations of the
  limb-darkened finite-disc correction throughout the wind, we find
  only a modest $\approx 10 \%$ higher mass-loss rate than in
  comparable models that assume a uniformly bright stellar disc.

As demonstrated below, this small effect can be readily
understood analytically, even though the complex spatial and
velocity dependence of the finite-disc correction factor generally
prohibits full analytic solutions for the steady-state equation of
motion. However, for the uniformly bright stellar-disc case, full
numerical solutions \citep[e.g.,][]{Friend86, Pauldrach86} show a
reduced mass-loss by a factor $\approx f(R_\star)^{1/\alpha} =
1/(1+\alpha)^{1/\alpha}$ as compared to the original CAK point star
rate,
\begin{eqnarray} 
  \dot{M} &\approx& \dot{M}_{\rm CAK} f(R_\star)^{1/\alpha} \nonumber \\
  &=& \frac{L}{c^2} \frac{\alpha}{1-\alpha} f(R_\star)^{1/\alpha} \left(
  \frac{\bar{Q} \Gamma_{\rm e}}{1-\Gamma_{\rm e}}
  \right)^{\frac{1}{\alpha}-1}
  \label{Eq:Mdot_cak}
\end{eqnarray} 
where a line-strength cut-off $\tau_{\rm max} \gg 1$ has been
assumed. 

We may understand this effect of the finite extent of the
stellar disc on the mass-loss rate by noting that, since $f(r)$
increases outward, the stellar surface now represents a ``critical
point'' from which it is hardest to accelerate the wind, thereby also
setting the maximal allowed mass-loss rate. Thus the simple
substitution $f \rightarrow f^{\rm ld}$ shows that including
photospheric limb-darkening increases the finite-disc corrected
mass-loss rate by $\approx 10 \%$ for reasonable values $\alpha
\approx 2/3$ \citep[see also][]{Cranmer95, Cure12}, as also
found above for the numerical model.

%\bibliographystyle{mn2e}
%\bibliography{my_refs}
\bibliography{sundqvist_fcl}

\end{document}